\documentclass[aps,prl,twocolumn,showpacs,superscriptaddress]{revtex4}
\usepackage{amsmath,amsfonts,amssymb}
\usepackage{color,graphicx,hyperref,epstopdf}
\usepackage{pdfpages}
\usepackage[paper=a4paper,margin=20mm]{geometry}
\begin{document}
\title{Unified nonclassicality criteria}

\author{S. Ryl}\email{sergej.ryl@uni-rostock.de}\affiliation{Arbeitsgruppe Theoretische Quantenoptik, Institut f\"ur Physik, Universit\"at Rostock, D-18051 Rostock, Germany}
\author{J. Sperling}\affiliation{Arbeitsgruppe Theoretische Quantenoptik, Institut f\"ur Physik, Universit\"at Rostock, D-18051 Rostock, Germany}
\author{E. Agudelo}\affiliation{Arbeitsgruppe Theoretische Quantenoptik, Institut f\"ur Physik, Universit\"at Rostock, D-18051 Rostock, Germany}
\author{M. Mraz}\affiliation{Arbeitsgruppe Experimentelle Quantenoptik, Institut f\"ur Physik, Universit\"at Rostock, D-18051 Rostock, Germany}
\author{S. K\"ohnke}\affiliation{Arbeitsgruppe Experimentelle Quantenoptik, Institut f\"ur Physik, Universit\"at Rostock, D-18051 Rostock, Germany}
\author{B. Hage}\affiliation{Arbeitsgruppe Experimentelle Quantenoptik, Institut f\"ur Physik, Universit\"at Rostock, D-18051 Rostock, Germany}
\author{W. Vogel}\affiliation{Arbeitsgruppe Theoretische Quantenoptik, Institut f\"ur Physik, Universit\"at Rostock, D-18051 Rostock, Germany}

\date{\today}

\begin{abstract}
	In this work we generalize the Bochner criterion addressing the characteristic function, i.e., the Fourier transform, of the Glauber-Sudarshan phase-space function.
	For this purpose we extend the Bochner theorem by including derivatives of the characteristic function.
	The resulting necessary and sufficient nonclassicality criteria unify previously known moment-based criteria with those based on the characteristic function.
	For applications of the generalized nonclassicality probes, we provide direct sampling formulas for balanced homodyne detection.
	A squeezed vacuum state is experimentally realized and characterized with our method.
	This complete framework---theoretical unification, sampling approach, and experimental implementation---presents an efficient toolbox to characterize quantum states of light for applications in quantum technology.
\end{abstract}

\pacs{42.50.Ct, 42.50.Dv, 03.65.Wj, 02.30.Nw}
\maketitle
\paragraph*{Introduction.}
	Quantum physics differs in various ways from a classical description of nature.
	Exploiting these differences became a major research interest for the aim of developing quantum technologies.
	One principle scope of this field is the formulation of measurable conditions, which are fulfilled for classical systems but may be violated for nonclassical systems.

	In quantum optics, a fundamental approach for separating classical light fields from quantum ones is defined in terms of the Glauber-Sudarshan $P$~function~\cite{Glauber1963,Sudarshan1963}.
	If a harmonic oscillator system can be solely described by coherent states and classical statistics, then its $P$~function is a classical probability density.
	Whenever the $P$~function does not exhibit such classical behavior, it identifies a nonclassical quantum state~\cite{Mandel1986}.
	In general, the $P$~function can represent a quasiprobability.
	For example, squeezed light cannot be characterized by a classical $P$~function, as has been experimentally confirmed~\cite{Kiesel2011,Agudelo2014}.

	In order to certify quantum features of states, a number of nonclassicality criteria have been formulated.
	Here, let us focus on two hierarchies of criteria which have been successfully applied in theory and experiment to determine the quantum character of light.
	Both of them prove the nonpositivity of the $P$~function in terms of experimentally accessible quantities.

	The first hierarchy is based on Bochner's theorem and the characteristic function (CF), being the Fourier transform of the $P$~function~\cite{Vogel2000,Richter2002}.
	Bochner's theorem expresses necessary and sufficient conditions for a function to be the Fourier transform of a classical probability density~\cite{Bochner1933}.
	It presupposes only the fact of the function to be defined for all arguments, so that it is nonlocal in phase space.
	The CF can be directly sampled from balanced homodyne detection (BHD) data, as shown in experiments~\cite{Lvovsky2002,Zavatta2007,Kiesel2009,Mari2011}, uncovering nonclassical phenomena through violations of the Bochner conditions.
	From the hierarchy of the latter, only the second-order condition has a clear geometrical interpretation.
	Higher-order conditions, however, fail to exhibit such a geometric meaning, since they correlate a higher number of points in phase space.

	The second hierarchy is formulated in terms of the so-called matrix of moments (MOM)~\cite{Shchukin2005a,Shchukin2005}, containing the statistical moments of the $P$~function.
	Some second-order minors of MOM are known to identify fundamental quantum effects, such as sub-Poisson~\cite{Mandel1979} or squeezed light~\cite{Slusher1985}.
	The MOM, however, includes more general quantum effects in terms of higher-order moments (cf., e.g.,~\cite{Hong1985,Hillery1987,Agarval1993,Klyshko1996}).
	Generalizations of this method even allow one to identify entanglement~\cite{Shchukin2005b,Miranowicz2010} and space-time-dependent quantum correlations~\cite{Vogel2008}.

	In the present contribution we combine the advantages of the CF and the MOM of the $P$~function resulting in a generalization of Bochner's theorem.
	This leads to a hierarchy of nonclassicality conditions embedding CF and MOM nonclassicality probes.
	Beyond the established cases, we obtain previously unknown conditions, which are shown to outperform the previously known criteria for specific mixed quantum states.
	In order to apply the proposed technique, we derive sampling formulas to reconstruct derivatives of the CF from BHD data.
	The path---starting with the theoretical treatment of unified nonclassicality probes---is finalized with our experimental implementation of the criteria for a squeezed state.
	Even though such states have a demanding sampling behavior, due to the exponential growth of their CF, the nonclassicality of the measured state is confirmed with a high significance.

\paragraph{Identification of nonclassical CFs.}
	Any single-mode quantum state of light can be given in Glauber-Sudarshan representation~\cite{Glauber1963,Sudarshan1963},
	\begin{equation}
		\hat{\rho}=\int d^2\alpha\, P(\alpha) |\alpha\rangle\langle\alpha|,
	\end{equation}
	where $P$ is a quasiprobability distribution.
	On the one hand, if  $P$ is a classical probability density, the state $\hat\rho$ has a classical counterpart.
	On the other hand, nonclassical states are characterized through negativities in the quasiprobability distribution, such as Fock or squeezed states.
	The $P$ function is often strongly singular~\cite{Sudarshan1963} and, hence, not always experimentally accessible.

	Even if the $P$~function is a highly nontrivial distribution, its CF,
	\begin{equation}\label{Eq:CharFct}
		\Phi(\beta)=\langle{:}e^{\beta\hat a^\dagger-\beta^\ast\hat a}{:}\rangle=\int d^2\alpha\, P(\alpha)e^{\beta\alpha^\ast-\beta^\ast\alpha},
	\end{equation}
	is a well-behaved function.
	Here ${:}\,\cdot\,{:}$ denotes the normal ordering prescription, and $\hat a$ and $\hat a^\dagger$ are the annihilation and creation operators, respectively.
	Bochner's conditions~\cite{Bochner1933}---for $\Phi(\beta)$ to be the CF of a classical probability distribution---were reformulated for quantum optics in~\cite{Richter2002}.
	Namely, $\Phi(\beta)$ is the CF of a classical state if the following conditions are fulfilled:
	(i) normalization, $\Phi(0)=1$;
	(ii) hermiticity, $\Phi(-\beta)=\Phi^\ast(\beta)$ for all $\beta\in\mathbb C$; and 
	(iii) positive semidefiniteness: for any positive integer $N$ and arbitrary complex numbers $\beta_1,\dots,\beta_N$ holds
	\begin{equation}\label{Eq:BochnerCondIII}
		\Phi=[\Phi(\beta_i-\beta_j)]_{i,j=1}^{N}\geq0,
	\end{equation}
	i.e., the matrix $\Phi$ is always positive semidefinite.
	Since conditions (i) and (ii) are valid for any state, the violation of condition (iii) discerns classical and nonclassical states.

	Employing Sylvester's theorem, the non-negativity in~\eqref{Eq:BochnerCondIII} can be written in terms of minors.
	For example, the second-order nonclassicality condition ($N=2$) reads as
	\begin{equation}
	 \det\Phi=1-|\Phi(\beta)|^2 <0,
	 \label{eq:loBocherviol}
	\end{equation}
	having a clear geometric interpretation.
	Whenever the absolute value of the CF exceeds 1, the state is nonclassical~\cite{Vogel2000}.
	Beyond this second-order minor, there is no clear geometric view of the higher-order conditions.

\paragraph*{Generalizing Bochner's theorem.}
	For the desired generalization, let us recall that for any classical state and any function $f(\alpha)$ holds
	\begin{equation}\label{Eq:NonNegCond1}
		\int d^2\alpha\, P(\alpha)|f(\alpha)|^2\geq 0.
	\end{equation}
	The relation of the inner product of two functions $g(\alpha)$ and $h(\alpha)$ to the inner product of their two-dimen\-sional Fourier transformed functions $\tilde g(\beta)$ and $\tilde h(\beta)$ is given from Parseval's theorem:
	$\pi^2\int d^2\alpha\, g^\ast(\alpha)h(\alpha)=\int d^2\beta\, \tilde g^\ast(\beta)\tilde h(\beta)$.
	Hence, the non-negativity condition~\eqref{Eq:NonNegCond1} can be written as
	\begin{equation}\label{Eq:NonNegCond2}
		\int d^2\beta\,\Phi(\beta)\mathrm{Aut}[\tilde f](\beta)\geq 0,
	\end{equation}
	using the Fourier transform of the function $|f(\alpha)|^2$ which is given by the autocorrelation function
		$\mathrm{Aut}[\tilde f](\beta)=\int d^2\gamma \tilde f^\ast(\gamma)\tilde f(\gamma+\beta)$.
	Choosing arbitrary non-negative integers $m_i$ and $n_i$ as well as complex numbers $\beta_i$ and $f_i$, we can define a function
	\begin{equation}
		\tilde f(\beta)=\sum_{i=1}^{N}f_i\partial_{\beta}^{n_i}\partial_{\beta^\ast}^{m_i}\delta(\beta-\beta_i).
	\end{equation}
	Now, the classicality condition~\eqref{Eq:NonNegCond2} reads as
	\begin{equation}\label{Eq:NonNegCond3}
		\sum_{i,j=1}^Nf_{i}f_{j}^{\ast}(-1)^{n_i+m_i}\partial_{\beta}^{n_i+m_j}\partial_{\beta^{\ast}}^{n_j+m_i}\Phi(\beta)\Big|_{\beta=\beta_i-\beta_j}\geq 0,
	\end{equation}
	(see the Supplemental Material~\cite{Supplement} for details).

	The definition of a generalized Bochner matrix~(GBM),
	\begin{equation}\label{Eq:BochnerMatrix}
		\partial\Phi=\left[(-1)^{n_i+m_i}\partial_{\beta}^{n_i+m_j}\partial_{\beta^{\ast}}^{n_j+m_i}\Phi(\beta)\Big|_{\beta=\beta_i-\beta_j}\right]_{i,j=1}^N,
	\end{equation}
	and the vector $\vec f=(f_1,\dots,f_N)^{\rm T}$ leads to the generalized version of Bochner's theorem.
	Before formulating the details, let us comment on some properties of $\partial \Phi$. 
	First, the GBM is Hermitian~\cite{Supplement}.
	Second, inequality~\eqref{Eq:NonNegCond3} may be formulated in the compact form $\vec f^{\,\dagger}\partial\Phi\vec f\geq 0$.
	Third, the elements of GBM can be also given as
	\begin{align}
		\nonumber
		&(-1)^{n_i+m_i}\partial_{\beta}^{n_i+m_j}\partial_{\beta^{\ast}}^{n_j+m_i}\Phi(\beta)\Big|_{\beta=\beta_i-\beta_j}\\
		=&(-1)^{n_i+n_j}\langle{:}
			\hat a^{\dagger}{}^{n_i+m_j}\hat a{}^{n_j+m_i}
			e^{[\beta_i-\beta_j]\hat a^\dagger-[\beta_i-\beta_j]^\ast\hat a}
		{:}\rangle.\label{Eq:MomentFormGBM}
	\end{align}

	{\it Theorem.}
	For any classical state it holds that the GBM is positive semidefinite, i.e., $\partial\Phi\geq 0$.\hfill$\blacksquare$

\paragraph*{Properties of the generalized theorem.}
	As mentioned earlier, a minor representation for the generalized Bochner theorem can be formulated.
	Namely, a state is nonclassical, if there exists a positive integer $N$, non-negative integers $\vec n=(n_1,\dots,n_N)^{\rm T}$ and $\vec m=(m_1,\dots,m_N)^{\rm T}$, and 
	complex numbers $\vec \beta=(\beta_1,\dots,\beta_N)^{\rm T}$ such that
	\begin{equation}\label{Eq:GBMcriterionDet}
		\det\partial\Phi<0.
	\end{equation}

	As an example, let us construct a nonobvious nonclassicality criterion.
	The parameters $N=3$, $\vec m=(0,1,0)^{\rm T}$, $\vec n=(0,0,1)^{\rm T}$, and $\vec \beta=(\beta,0,\beta)^{\rm T}$ for $\beta\in\mathbb C$ produce the GBM:
	\begin{align}\label{Eq:ExampleGBM}
		\partial\Phi=&
		\begin{pmatrix}
		\Phi(0)&\partial_\beta\Phi(\beta)&\partial_{\beta^\ast}\Phi(0)\\
		-\partial_{\beta^\ast}\Phi(-\beta)&-\partial_\beta\partial_{\beta^\ast}\Phi(0)&-\partial_{\beta^\ast}^2\Phi(-\beta)\\
		-\partial_{\beta}\Phi(0)&-\partial_{\beta}^2\Phi(\beta)&-\partial_\beta\partial_{\beta^\ast}\Phi(0)
		\end{pmatrix}.
	\end{align}
	This previously unknown nonclassicality criterion~\eqref{Eq:GBMcriterionDet} for the GBM~\eqref{Eq:ExampleGBM} combines moments and the CF of a nonclassical state.
	That is, for $\beta=0$, the minor corresponds to the quadrature squeezing condition~\cite{Supplement,Shchukin2005}
	\begin{equation}\label{Eq:squeezing}
		\det(\partial\Phi)\Big|_{\beta=0}{=}\frac14\langle{:}[\Delta\hat{x}(\varphi_{\min})]^2{:}\rangle\langle{:}[\Delta\hat{x}(\varphi_{\max})]^2{:}\rangle{<}0,
	\end{equation}
	using the relation~\eqref{Eq:MomentFormGBM}.
	The general relation between CF and MOM nonclassicality conditions will be derived in the following by applying the above theorem.

	First, we may compare the GBM, Eq.~\eqref{Eq:BochnerMatrix}, with the corresponding MOM and Bochner matrices of the same dimensionality $N$.
	A Bochner matrix $\Phi$, defined in Eq.~\eqref{Eq:BochnerCondIII}, corresponds to a GBM $\partial\Phi$ in Eq.~\eqref{Eq:BochnerMatrix}, if we set the integers as $m_1=\dots=m_N=n_1=\dots=n_N=0$.
	Thus, the matrix elements of the GBM reduce to
	\begin{equation}
		(-1)^{0}\partial_{\beta}^{0}\partial_{\beta^{\ast}}^{0}\Phi(\beta)\Big|_{\beta=\beta_i-\beta_j}=\Phi(\beta_i-\beta_j).
	\end{equation}
	Since constraints (i) and (ii) of a CF are already fulfilled, this special case of the non-negativity of the GBM is equivalent to condition (iii) in inequality~\eqref{Eq:BochnerCondIII}.
	Consequently, $\partial\Phi\geq0$ implies $\Phi\geq 0$, which is already necessary and sufficient to characterize a classical state~\cite{Richter2002}.
	Note that the Bochner criterion is nonlocal in phase space, since it correlates distant points $\beta_i$ and $\beta_j$ ($i\neq j$).

	Second, a MOM corresponds to a GBM, if we choose $\beta_1=\dots=\beta_N=0$ in~\eqref{Eq:MomentFormGBM}, since the matrix elements simplify to $\partial\Phi=[(-1)^{n_i+n_j}\langle \hat a^\dagger{}^{n_i+m_j}\hat a^{n_j+m_i}\rangle]_{i,j=1}^N,$.
	Using the representation of the non-negativity in Eq.~\eqref{Eq:NonNegCond3}, we obtain
	\begin{align}
		\sum_{i,j=1}^Nf'_if'_j{}^\ast\langle \hat a^\dagger{}^{n_i+m_j}\hat a^{n_j+m_i}\rangle\geq 0,
	\end{align}
	with $f'_k=(-1)^{n_k} f_k$ for $k=1,\dots,N$.
	In this special case, the generalized Bochner theorem coincides with a MOM criterion for annihilation and creation operators~\cite{Shchukin2005}.
	Since the MOM is represented through derivatives of the CF at the origin, $\beta=0$, they are local characteristics in phase space.
	
	The previously independent approaches of the CF and the MOM have been unified and generalized by our method.
	Nonlocal Bochner and local MOM nonclassicality probes occur simultaneously in the nonclassicality theorem based on the GBM.
	In the following, we provide an example of a particular GBM inferring quantumness of a state which is inaccessible via the corresponding MOM and Bochner approaches.
	
\begin{figure}[t]
	  \includegraphics[width=85mm]{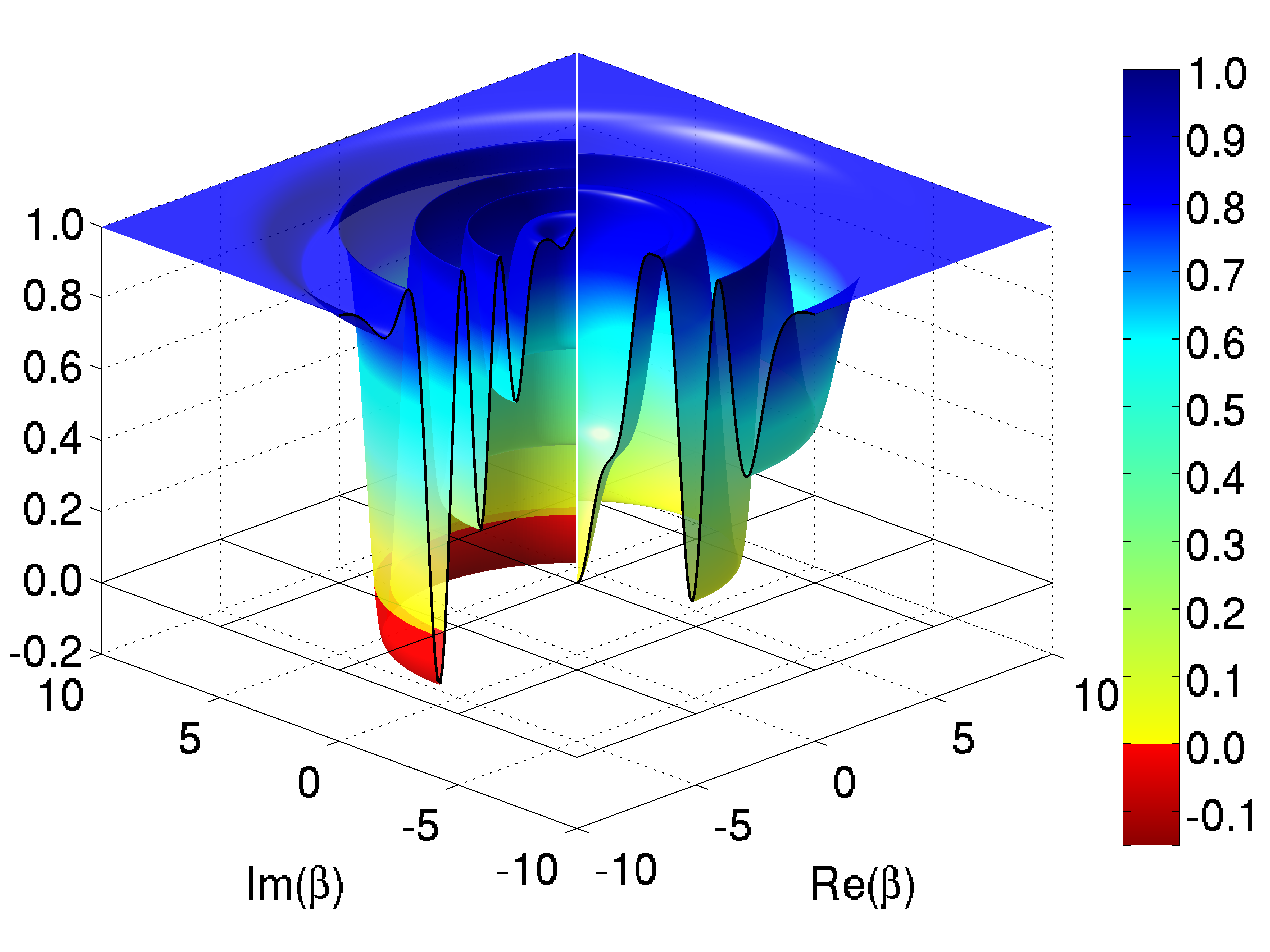}
	  \caption{
		(Color online)
		Left half: The scaled ($3.06\times$) minor~\eqref{Eq:GBMSRex} for the state~\eqref{Eq:TheorExample} is shown for the parameters $p_0=0.944,~\bar n_0=0.1,~p_3=0.03,~\bar n_3=0.12,~p_4=0.026,~\bar n_4=0.182$.
		The displayed negativity at $|\beta|\approx5.8$ of this determinant verifies the nonclassicality of the state under study.
		Right half: The lowest-order criterion~(\ref{eq:loBocherviol}) does not verify the nonclassical nature of the state.
		\label{Fig:ThExample}}
	\end{figure}

	Let us consider the situation, when the condition~\eqref{eq:loBocherviol} does not verify the nonclassicality of a given state.
	Additionally, the MOM minor
	\begin{equation}
		\det\begin{pmatrix}1&\langle\hat{a}\rangle\\\langle\hat{a}^\dag\rangle&\langle{:}\hat{a}^\dag\hat{a}{:}\rangle\end{pmatrix}\geq 0,\label{Eq:MoMminor}
	\end{equation}
	is non-negative for any state, as it represents the mean number of incoherent photons~\cite{Shchukin2005}.
	The second-order minor resulting from the GBM and the choice $m_1=m_2=n_1=0$, $n_2=1$ and $\beta=\beta_1-\beta_2$ reads as:
	\begin{align}
		\det\partial\Phi=&\nonumber
		\det\left(\begin{array}{cc}1&-\langle{:} \hat{a}e^{\beta\hat a^\dagger-\beta^\ast\hat a}{:}\rangle\\-\langle{:}\hat{a}^\dag e^{-\beta\hat a^\dagger+\beta^\ast\hat a}{:}\rangle&\langle{:}\hat{a}^\dag\hat{a}{:}\rangle \end{array}\right)
		\\=&\langle{:}\hat{a}^\dag\hat{a}{:}\rangle-\left|\langle{:}\hat{a}e^{\beta\hat a^\dagger-\beta^\ast\hat a}{:}\rangle\right|^2=\langle [\Delta\hat A^\dagger][\Delta\hat A]\rangle,
		\label{Eq:GBMSRex}
	\end{align}
	with $\hat A=\partial_{\beta^\ast}e^{\beta\hat a^\dagger-\beta^\ast\hat a}$.
	Now, we construct a mixed state, $\hat\rho_{\rm mix}$, being a mixture of a thermal state with three- and four-photon-added thermal states~\cite{Agarval1992,Parigi2008,Kiesel2008}.
	Its CF is
	\begin{align}
		\Phi_{\rm mix}(\beta)&=p_0 e^{-\bar{n}_0|\beta|^2}+p_3 {L}_3[(1+\bar{n}_3)|\beta|^2]e^{-\bar{n}_3|\beta|^2}\nonumber\\
		&+p_4{L}_4[(1+\bar{n}_4)|\beta|^2]e^{-\bar{n}_4|\beta|^2},\label{Eq:TheorExample}
	\end{align}
	with probabilities $p_i$, mean numbers of thermal photons $\bar n_i$, and the Laguerre polynomials ${L}_i(z)$ ($i=0,3,4$).
	The visualization of the minor~\eqref{Eq:GBMSRex} for this state is given in Fig.~\ref{Fig:ThExample}.
	It can be seen that negativities infer the quantumness of this state, which can neither be observed through the nonclassicality condition~\eqref{eq:loBocherviol} nor the MOM minor~\eqref{Eq:MoMminor} (see~\cite{Supplement} for details).

\paragraph*{Sampling the CF and its derivatives.}
	Properties of quantum states of light may be directly sampled from BHD~\cite{Welsch1999,Lvovski2009}.
	In this setup one measures the statistics of the field quadrature operator, $\hat x(\varphi)=e^{i\varphi}\hat a+e^{-i\varphi}\hat a^\dagger$, resulting in $M$ data points of quadrature values $[x_j(\varphi)]_{j=1}^M$ for a fixed phase $\varphi$.
	The CF~\eqref{Eq:CharFct} is readily obtained from the set of data,
	\begin{equation}
	 \Phi(\beta)\approx e^{|\beta|^2/2}\frac{1}{M}\sum_{j=1}^Me^{i|\beta|x_j(\varphi)},
	\label{eq:samplePhi}
	\end{equation}
	where $\arg\beta=\pi/2-\varphi$ (cf.~\cite{Lvovsky2002,Zavatta2007,Kiesel2009}).
	For sampling the derivatives in an analogous way, the knowledge of two noncommuting operators, $[\hat{a},\hat{a}^\dag]=\hat 1$, is required.
	Hence, another ansatz for reconstructing the GBM elements has to be developed.

	In order to apply our generalized Bochner conditions to experimental data, we need to reconstruct the characteristic function and its derivatives.
	An alternative way to reconstruct the CF is based on its Taylor expansion,
	\begin{equation}\label{Eq:Taylor}
		\Phi(\beta)=\langle:{\rm e}^{\beta\hat{a}^\dag-\beta^\ast\hat{a}}:\rangle=\sum_{k,l=0}^{\infty}\frac{\beta^k}{k!}\frac{(-\beta^\ast)^l}{l!}\langle\hat{a}^{\dag k}\hat{a}^l\rangle.
	\end{equation}
	The normally ordered moments can be represented with quadrature distributions $p(x,\varphi)$ and Hermite polynomials ${H}_n(z)$ as~\cite{Richter1996},
	\begin{align}
		\langle \hat{a}^{\dag k}\hat{a}^l\rangle=\int_{0}^{\pi}d\varphi\int_{-\infty}^{+\infty} dx\,& \frac{p(x,\varphi)}{\pi}\\
		&\times\frac{e^{i(k-l)\varphi}k!l!}{\sqrt{2^{k+l}}(k+l)!}{ H}_{k+l}\left(\frac{x}{\sqrt{2}}\right).\nonumber
	\end{align}
	The double series~\eqref{Eq:Taylor} can be evaluated as
	\begin{align}
		\Phi(\beta)=\int_{0}^{\pi}d\varphi\int_{-\infty}^{+\infty} dx&\,\frac{p(x,\varphi)}{\pi(\gamma+\gamma^\ast)}\label{Eg:LimitOfDoubleSeries}\\
		&\times\left[\gamma e^{x\gamma-\gamma^2/2}+\gamma^\ast e^{-x\gamma^\ast-\gamma^{\ast2}/2}\right],\nonumber
	\end{align}
	with $\gamma=\beta e^{i\varphi}$  (see~\cite{Supplement,PrudnikovBuch} for more technical details).
	By differentiation of \eqref{Eg:LimitOfDoubleSeries} the following pattern function emerges:
	\begin{align}
		D_{q}^{r}(x,\gamma)=\sum_{\substack{k_1+k_2\\+k_3=r}}&\dfrac{r!(-1)^{q+k_1+k_3}(q+k_1)!2^{-k_3/2}}{k_1!k_2!k_3!(\gamma+\gamma^\ast)^{q+k_1+1}}x^{k_2}\nonumber\\
		&\times\exp(x\gamma-\gamma^2/2){H}_{k_3}(\gamma/\sqrt{2}),
	\end{align}
	and $D_q^r\equiv0$ for $q{<}0$ or $r{<}0$.
	Now, the derivatives of the CF can be directly sampled as
	\begin{align}\label{Eq:SamplingFormula}
		\partial_{\beta}^m&\partial_{\beta^\ast}^n\Phi(\beta)\approx\frac{1}{M}\sum_{j=1}^{M}e^{i(m-n)\varphi_{j}}\\
		\times&\left[mD_{n}^{m-1}(x_j,\beta e^{i\varphi_j})+\beta e^{i\varphi_j} D_{n}^m(x_j,\beta e^{i\varphi_j})\right.\nonumber\\
		+&\left.nD_{m}^{n-1}(-x_j,\beta^\ast e^{- i\varphi_j})+\beta^\ast e^{- i\varphi_j} D_{m}^n(-x_j,\beta^\ast e^{- i\varphi_j})\right],\nonumber
	\end{align}
	with the need of a uniform distribution of the measured data $(x_j,\varphi_j)_{j=1}^{M}$ with respect to the phase $\varphi$.

\begin{figure}[t]
		\centering
		\includegraphics*[width=75mm]{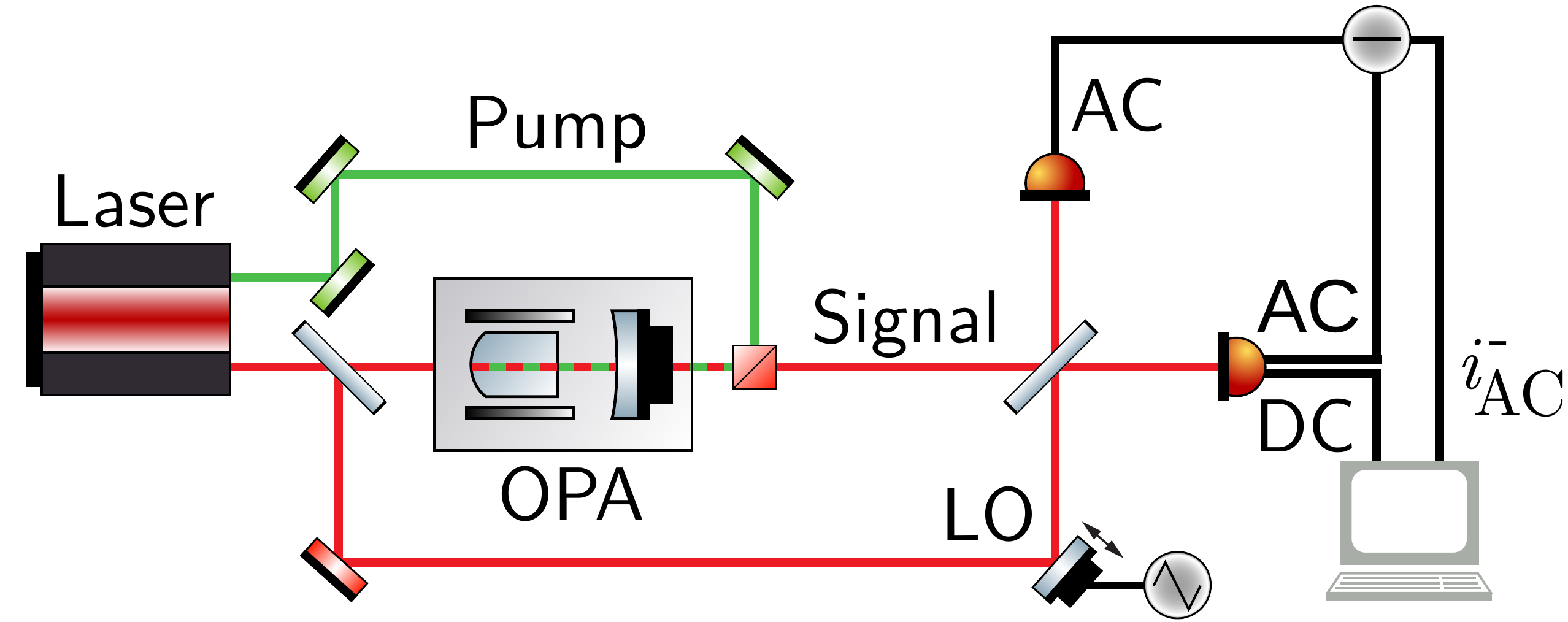}
		\caption{(Color online)
			Experimental scheme for the generation and measurement of a squeezed state.
			The state was generated in an OPA with an 7\%MgO:LiNbO$_{3}$ crystal, which was pumped with a 290\,\rm{mW} laser beam at 532\,\rm{nm} generating $-4.13\,$dB squeezing at 1064\,\rm{nm}.
			A BHD (LO power 1.23\,\rm{mW}) is set up for the detection.
			The phase of the LO was altered continuously by applying a triangular alternating voltage onto the piezo that is responsible for the mirror position.
		}\label{fig:ExpSetup}
	\end{figure}

\paragraph*{Experimental implementation.}
	We generated the squeezed field through a hemilithic, standing wave, nonlinear cavity (see Fig.~\ref{fig:ExpSetup}) serving as an optical parametric amplifier (OPA).
	As $\chi^{(2)}$ nonlinear medium we used an 8-mm-long, 7\% magnesium-oxide-doped lithium niobate (7\%MgO:LiNbO$_{3}$) crystal.
	The OPA is pumped with a strong laser beam (290\,\rm{mW}) at 532\,\rm{nm}, resulting in a classical gain of 3.3, which yields $-4.13\,{\rm dB}$ squeezing and 6.11\,dB anti-squeezing at 1064\,\rm{nm}.
	A BHD was implemented~\cite{Smithey1993,Breitenbach1995} with 98\% visibility and a quantum efficiency of 90\%.
	Along with additional losses in the squeezed field due to losses in optical components and an escape efficiency smaller than unity, this results in a overall efficiency of 77\%.
	The optical phase of the signal, relative to the local oscillator (LO), was controlled by a mirror mounted on a piezoactuator.
	A continuous variation of the optical phase ensures an equally distributed phase of the quadrature data points $(x_j,\varphi_j)_{j=1}^M$~\cite{Agudelo2014}.

\begin{figure}[ht]
		{sampled minor $\det(\partial\Phi)$}\vspace*{-0.1cm}
		\includegraphics[width=8.5cm]{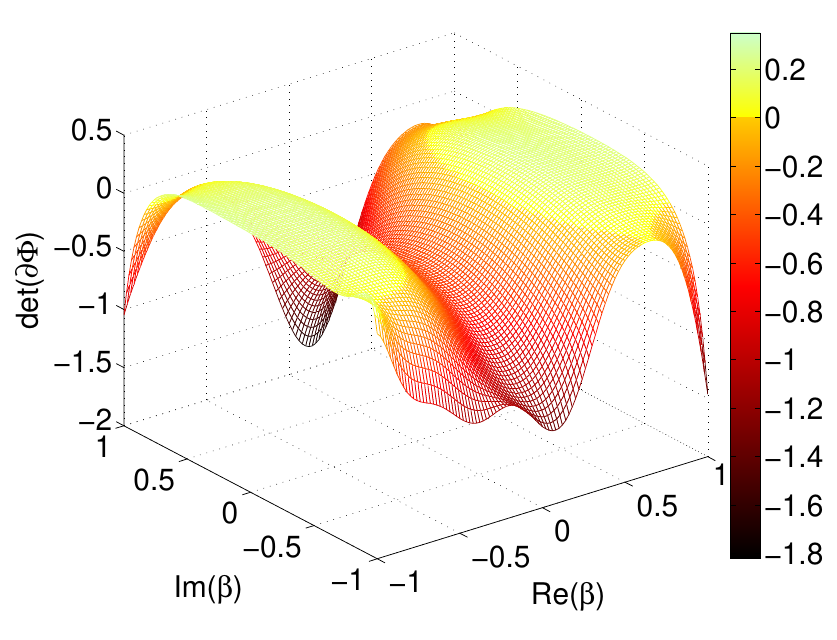}
		{signed significance $\Sigma$}\vspace*{-0.2cm}
		\includegraphics[width=7cm]{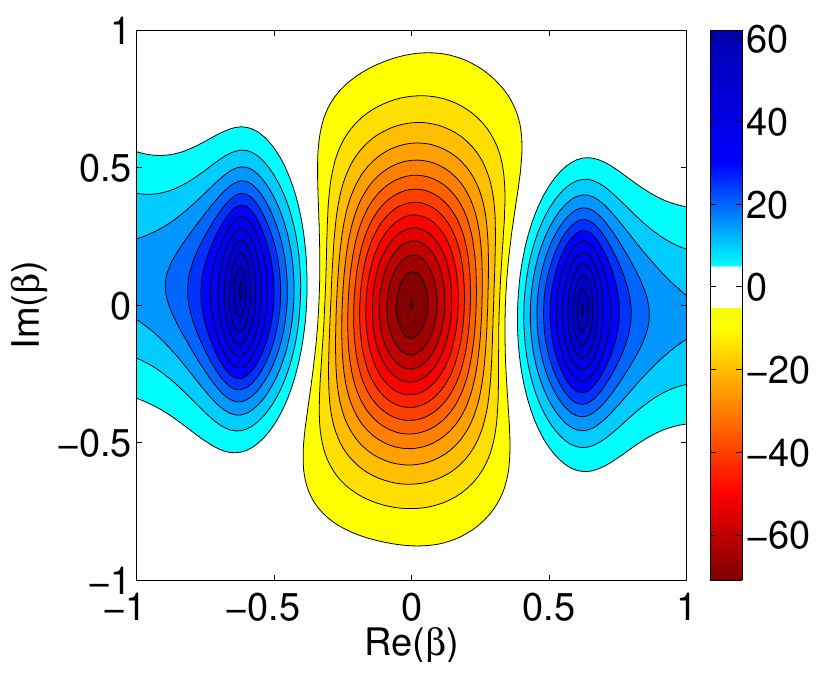}
		\caption{(Color online)
 			The top plot depicts the experimental implementation of the nonclassicality condition for the determinant of~\eqref{Eq:ExampleGBM}, $\det(\partial\Phi)<0$.
 			Negative values demonstrate that the GBM criterion is capable to prove the nonclassicality of the generated state.
 			The bottom plot shows the signed significance $\Sigma[\det(\partial\Phi)]=\det(\partial\Phi)/\sigma[\det(\partial\Phi)]$ up to $-70$ standard deviations at the origin and $+60$ in side peaks.
 			The region with $|\Sigma|<5$ is shown white.
 			The error $\sigma[\det(\partial\Phi)]$ is obtained via a linear error propagation of the GBM elements.
 		}\label{Fig:Reconstruct}
		
	\end{figure}

	Using the sampling formula~\eqref{Eq:SamplingFormula}, we reconstructed the determinant of the GBM~\eqref{Eq:ExampleGBM}.
	It is important to mention that the CF of the squeezed state is diverging in one direction of $\beta$, which causes high sampling errors.
	This fact makes such a state an optimal test for demonstrating the applicability of our sampling approach under difficult premises.
	Figure~\ref{Fig:Reconstruct} shows the determinant $\det(\partial\Phi)$ of the GBM, Eq.~\eqref{Eq:ExampleGBM}, together with its signed significance,
	\begin{equation}
	 \Sigma[\det(\partial\Phi)]=\frac{\det(\partial\Phi)}{\sigma[\det(\partial\Phi)]},
	 \label{eq:signedSign}
	\end{equation}
	where $\sigma[\det(\partial\Phi)]$ denotes the standard deviation.
	Without an additional search for optimal phases, the squeezing condition~\eqref{Eq:squeezing} is fulfilled as $\det(\partial\Phi)|_{\beta=0}=-0.469\pm0.007$.
	At this point the highest significance of a negative value of above 70 standard deviations is reached.
	From the decay of the significances, it can be observed that the squeezed and anti-squeezed axes are---up to a slight rotation---the imaginary and real axes, respectively.

\paragraph*{Conclusions.}\label{sec:conclusions}
	We derived necessary and sufficient nonclassicality probes by generalizing Bochner's theorem.
	This generalization has been proven to embed two hierarchies of previously known nonclassicality criteria.
	Namely, the conditions resulting from the original Bochner's theorem and the matrix of moments can be reproduced by suitable parameter choice.
	Our generalized Bochner theorem makes use of the advantages of the original Bochner theorem, being nonlocal characteristics in phase space, and of the matrix of moments, directly yielding quantum features such as squeezing.

	We explicitly constructed a nonclassical state, for which specific second-order moments and Bochner criteria fail, but the corresponding generalized second-order criterion truly visualizes its quantumness.
	Pattern functions have been derived to sample our criteria from balanced homodyne detection data.
	For a direct demonstration, a squeezed state---having a demanding sampling behavior---has been prepared and measured.
	We could successfully demonstrate that our approach certifies the nonclassicality of this state with high significance.

	Thus, we explored the entire process from formulating a unified technique to uncover quantumness until its successful experimental implementation within this work.
	Moreover, it is straightforward to generalize our method to multimode scenarios by applying the present techniques.
	The here established method allows one to infer quantum effects of previously unknown structures in phase space beyond local properties, such as higher-order moments, and nonlocal properties of the characteristic function, as addressed by the original form of the Bochner theorem.
	Eventually, this unified characterization of quantum features will allow one to assist quantum technologies with properly designed quantum states of light.

\paragraph*{Acknowledgments.}
	The authors are grateful to B.~K\"uhn for helpful discussions.
	This work is supported by the Deutsche Forschungs\-gemeinschaft through SFB~652.

\widetext\newpage
\newgeometry{margin=0mm}
	{\label{RylSupplement20150429}

\section*{Supplementary material}

# Supplemental Material
# Unifying Nonclassicality Criteria by Generalizing Bochner's Theorem

S. Ryl,[1,*] J. Sperling,[1] E. Agudelo,[1] M. Mraz,[2] S. Köhnke,[2] B. Hage,[2] and W. Vogel[1]

[1]*Arbeitsgruppe Theoretische Quantenoptik, Institut für Physik, Universität Rostock, D-18051 Rostock, Germany*
[2]*Arbeitsgruppe Experimentelle Quantenoptik, Institut für Physik, Universität Rostock, D-18051 Rostock, Germany*
(Dated: April 29, 2015)

Section I includes details on the generalized Bochner matrix. In Section II we discuss the physical meaning of some low-order minors. We also highlight properties of special kinds of states. In Section III the detailed derivation of sampling formulas and the proof of pattern functions to be analytical is given. Additional information about the experimentally realized squeezed state and the condition used to prove its quantumness is shown in Section IV.

We use the abbreviations of the main part: generalized Bochner matrix (GBM), characteristic function (CF), and matrix of moments (MoM).

## I. THE GENERALIZED BOCHNER MATRIX

In this section the GBM is derived and its properties are studied. The special cases of the Bochner matrix and the MoM are retrieved.

The autocorrelation function of a function $\tilde{f}(\beta) = \sum_{i=1}^{N} f_i \partial_\beta^{n_i} \partial_{\beta^*}^{m_i} \delta(\beta - \beta_i)$ reads as

$$\mathrm{Aut}[\tilde{f}](\beta) = \int d^2\gamma\, \tilde{f}(\beta + \gamma) \tilde{f}^*(\gamma) = \sum_{i,j=1}^{N} f_i f_j^* \int d^2\gamma\, \left[\partial_\gamma^{n_i} \partial_{\gamma^*}^{m_i} \delta(\beta + \gamma - \beta_i)\right] \left[\partial_\gamma^{m_j} \partial_{\gamma^*}^{n_j} \delta(\gamma - \beta_j)\right]. \tag{1}$$

Using integration by parts we get

$$\mathrm{Aut}[\tilde{f}](\beta) = \sum_{i,j=1}^{N} f_i f_j^* (-1)^{n_j + m_j} \int d^2\gamma\, \delta(\gamma - \beta_j) \partial_\gamma^{n_i + m_j} \partial_{\gamma^*}^{m_i + n_j} \delta(\beta + \gamma - \beta_i)$$

$$= \sum_{i,j=1}^{N} f_i f_j^* (-1)^{n_j + m_j} \partial_\beta^{n_i + m_j} \partial_{\beta^*}^{m_i + n_j} \delta(\beta + \beta_j - \beta_i). \tag{2}$$

The evaluation of the inner product $I$ of this autocorrelation function with the CF $\Phi(\beta)$ yields

$$I = \int d^2\beta\, \Phi(\beta) \mathrm{Aut}[\tilde{f}](\beta) = \sum_{i,j=1}^{N} f_i f_j^* (-1)^{n_j + m_j} \int d^2\beta\, \Phi(\beta) \partial_\beta^{n_i + m_j} \partial_{\beta^*}^{m_i + n_j} \delta(\beta + \beta_j - \beta_i) \tag{3}$$

$$= \sum_{i,j=1}^{N} f_i f_j^* (-1)^{n_i + m_i} \int d^2\beta\, \delta(\beta + \beta_j - \beta_i) \partial_\beta^{n_i + m_j} \partial_{\beta^*}^{m_i + n_j} \Phi(\beta) = \sum_{i,j=1}^{N} f_i f_j^* (-1)^{n_i + m_i} \left. \partial_\beta^{n_i + m_j} \partial_{\beta^*}^{m_i + n_j} \Phi(\beta) \right|_{\beta = \beta_i - \beta_j}.$$

Defining the vector $\vec{f} = (f_1, \ldots, f_N)^{\mathrm{T}}$ and the GBM

$$\partial\Phi = \begin{pmatrix} \partial\Phi_{1,1} & \cdots & \partial\Phi_{1,N} \\ \vdots & \ddots & \vdots \\ \partial\Phi_{N,1} & \cdots & \partial\Phi_{N,N} \end{pmatrix} = \left( (-1)^{n_i + m_i} \left. \partial_\beta^{n_i + m_j} \partial_{\beta^*}^{m_i + n_j} \Phi(\beta) \right|_{\beta = \beta_i - \beta_j} \right)_{i,j=1}^{N}, \tag{4}$$

we can write the inner product $I$ as

$$I = \vec{f}^{\,\dagger}\, \partial\Phi\, \vec{f}. \tag{5}$$

---

*Electronic address: sergej.ryl@uni-rostock.de

The elements of the GBM can be also computed as

$$\partial \Phi_{i,j} = (-1)^{n_i+m_i} \partial_\beta^{n_i+m_j} \partial_{\beta^*}^{m_i+n_j} \langle :e^{\beta \hat{a}^\dagger - \beta^* \hat{a}}: \rangle \Big|_{\beta=\beta_i-\beta_j} = (-1)^{n_i+n_j} \langle :\hat{a}^{\dagger n_i+m_j} \hat{a}^{m_i+n_j} e^{[\beta_i-\beta_j]\hat{a}^\dagger - [\beta_i-\beta_j]^* \hat{a}}: \rangle. \quad (6)$$

The GBM is Hermitian, because for all $i, j$ holds

$$\begin{aligned}
\partial \Phi_{j,i}^* =& (-1)^{n_j+n_i} \langle :\hat{a}^{n_j+m_i} \hat{a}^{\dagger m_j+n_i} e^{[\beta_j-\beta_i]^* \hat{a} - [\beta_j-\beta_i]\hat{a}^\dagger}: \rangle \\
=& (-1)^{n_i+n_j} \langle :\hat{a}^{\dagger n_i+m_j} \hat{a}^{m_i+n_j} e^{[\beta_i-\beta_j]\hat{a}^\dagger - [\beta_i-\beta_j]^* \hat{a}}: \rangle = \partial \Phi_{i,j}.
\end{aligned} \quad (7)$$

For the special case $n_i = m_i = 0$ for all $i$, we get the corresponding Bochner matrix

$$\partial \Phi = \begin{pmatrix} \Phi(0) & \Phi(\beta_1 - \beta_2) & \cdots \\ \Phi(\beta_2 - \beta_1) & \Phi(0) & \cdots \\ \vdots & \vdots & \ddots \end{pmatrix}. \quad (8)$$

Note that $\operatorname{tr} \hat{\rho} = 1 = \int d^2\alpha\, P(\alpha) = \Phi(0)$ for any quantum state $\hat{\rho}$.

For relating the GBM to the corresponding MoM, one may generate a list of pairs of natural numbers:

$$\begin{pmatrix} 0 \\ 0 \end{pmatrix}, \begin{pmatrix} 1 \\ 0 \end{pmatrix}, \begin{pmatrix} 0 \\ 1 \end{pmatrix}, \begin{pmatrix} 2 \\ 0 \end{pmatrix}, \begin{pmatrix} 1 \\ 1 \end{pmatrix}, \begin{pmatrix} 0 \\ 2 \end{pmatrix}, \ldots, \begin{pmatrix} 0 \\ n-1 \end{pmatrix}, \begin{pmatrix} n \\ 0 \end{pmatrix}, \begin{pmatrix} n-1 \\ 1 \end{pmatrix}, \ldots, \begin{pmatrix} 1 \\ n-1 \end{pmatrix}, \begin{pmatrix} 0 \\ n \end{pmatrix}, \begin{pmatrix} n+1 \\ 0 \end{pmatrix}, \ldots.$$

The upper and lower vector components yields an ordering as

$$(n_1, n_2, \ldots) = (0, 1, 0, 2, 1, 0, \ldots) \quad \text{and} \quad (m_1, m_2, \ldots) = (0, 0, 1, 0, 1, 2, \ldots). \quad (9)$$

For $\beta_i = 0$ for all $i$, the GBM corresponds to the MoM [1],

$$\partial \Phi = D \begin{pmatrix} 1 & \langle \hat{a} \rangle & \langle \hat{a}^\dagger \rangle & \cdots \\ \langle \hat{a}^\dagger \rangle & \langle \hat{a}^\dagger \hat{a} \rangle & \langle \hat{a}^{\dagger 2} \rangle & \cdots \\ \langle \hat{a} \rangle & \langle \hat{a}^2 \rangle & \langle \hat{a}^\dagger \hat{a} \rangle & \cdots \\ \vdots & \vdots & \vdots & \ddots \end{pmatrix} D, \quad (10)$$

up to the unitary transformation $D = \operatorname{diag}\left((-1)^{n_i}\right)_{i \in \mathbb{N}\setminus\{0\}} = D^\dagger = D^{-1}$.

## II. LOW-ORDER MINORS AND SOME EXAMPLES

### A. Interpretation of low-order minors

Second and third order minors can be considered as the variance or cross-correlation, respectively. This means ($\Delta X = X - \langle X \rangle$):

$$\det \begin{pmatrix} 1 & \langle A \rangle \\ \langle A^* \rangle & \langle A^* A \rangle \end{pmatrix} = \langle A^* A \rangle - \langle A^* \rangle \langle A \rangle = \langle \Delta A^* \Delta A \rangle, \quad (11)$$

$$\det \begin{pmatrix} 1 & \langle A \rangle & \langle B \rangle \\ \langle A^* \rangle & \langle A^* A \rangle & \langle A^* B \rangle \\ \langle B^* \rangle & \langle B^* A \rangle & \langle B^* B \rangle \end{pmatrix} = \langle \Delta A^* \Delta A \rangle \langle \Delta B^* \Delta B \rangle - \langle \Delta A^* \Delta B \rangle \langle \Delta B^* \Delta A \rangle. \quad (12)$$

Second order nonclassicality probes of the Bochner criterion or a specific MoM criterion are

$$\det \begin{pmatrix} 1 & \Phi(\beta) \\ \Phi(-\beta) & 1 \end{pmatrix} = 1 - |\Phi(\beta)|^2 < 0, \quad (13)$$

$$\det \begin{pmatrix} 1 & \langle \hat{a} \rangle \\ \langle \hat{a}^\dagger \rangle & \langle \hat{a}^\dagger \hat{a} \rangle \end{pmatrix} = \langle [\Delta \hat{a}]^\dagger [\Delta \hat{a}] \rangle < 0. \quad (14)$$

## B. Multi-photon-added-thermal-states.

Note that states with a Fock diagonal statistical operator $\hat{\rho}$ have a rotationally symmetric CF, $\Phi(\beta) = \Phi(|\beta|)$. One class of such kind of states are multi-photon-added-thermal-states [2]. The $P$ function of a $m$ photon-added thermal state ($m$PATS) is

$$P(\alpha) = \frac{(-1)^m}{\bar{n}^m} L_m\left[\left(1 + \frac{1}{\bar{n}}\right)|\alpha|^2\right] \cdot \frac{1}{\pi\bar{n}} e^{-|\alpha|^2/\bar{n}}, \quad (15)$$

where $\bar{n}$ is the mean photon number of the thermal background and $L_m$ is the $m$th Laguerre polynomial. The latter can be expressed by the Rodrigues formula: $L_m(z) = \frac{1}{m!} e^z \frac{d^m}{dz^m}(z^m e^{-z})$. The Fourier transform of the $P$ function of the $m$PATS $\hat{\rho}_{m\text{PATS}}(\bar{n})$ is

$$\Phi(\beta) = L_m[(1+\bar{n})|\beta|^2] \cdot e^{-\bar{n}|\beta|^2}. \quad (16)$$

The single-photon-added-thermal-state (SPATS or, in our notion, 1PATS) has been experimentally realized in [3] and analyzed in [4]. We use a statistical mixture of a thermal state (0PATS), a 3PATS, and a 4PATS to show the advantage of the GBM approach. That is

$$\hat{\rho}_{\text{mix}} = p_0 \hat{\rho}_{0\text{PATS}}(\bar{n}_0) + p_3 \hat{\rho}_{3\text{PATS}}(\bar{n}_3) + p_4 \hat{\rho}_{4\text{PATS}}(\bar{n}_4), \quad (17)$$

with the characteristic function:

$$\Phi_{\text{mix}}(\beta) = p_0 e^{-\bar{n}_0 |\beta|^2} + p_3 L_3[(1+\bar{n}_3)|\beta|^2] e^{-\bar{n}_3|\beta|^2} + p_4 L_4[(1+\bar{n}_4)|\beta|^2] e^{-\bar{n}_4|\beta|^2}. \quad (18)$$

The chosen parameters are the probabilities $p_0 = 0.944, p_3 = 0.03, p_4 = 0.026$ and mean numbers of thermal photons $\bar{n}_0 = 0.1, \bar{n}_6 = 0.12, \bar{n}_9 = 0.182$. The large contribution of the classical thermal state, $p_0 \gg p_3, p_4$, does not allow to proof the nonclassicality of this state with Bochner condition in Eq. (13); see dashed line in Fig. 1. The MoM criterion in (14) reads as $\langle(\hat{a} - \langle\hat{a}\rangle)^\dagger(\hat{a} - \langle\hat{a}\rangle)\rangle$, which cannot be negative since $\hat{X}^\dagger\hat{X} \geq 0$ for all $\hat{X}$. However, by using the corresponding determinant of the GBM in the main body,

$$\det \partial\Phi = \det\begin{pmatrix} 1 & \langle:\partial^*_\beta \hat{a} e^{\beta \hat{a}^\dagger - \beta^* \hat{a}}:\rangle \\ \langle:\partial_\beta e^{-\beta \hat{a}^\dagger + \beta^* \hat{a}}:\rangle & \langle:\hat{a}^\dagger \hat{a}:\rangle \end{pmatrix} = \langle:\hat{a}^\dagger\hat{a}:\rangle - \left|\langle:\hat{a} e^{\beta\hat{a}^\dagger - \beta^*\hat{a}}:\rangle\right|^2, \quad (19)$$

we can prove the state (17) to be nonclassical; see the solid line in Fig. 1.

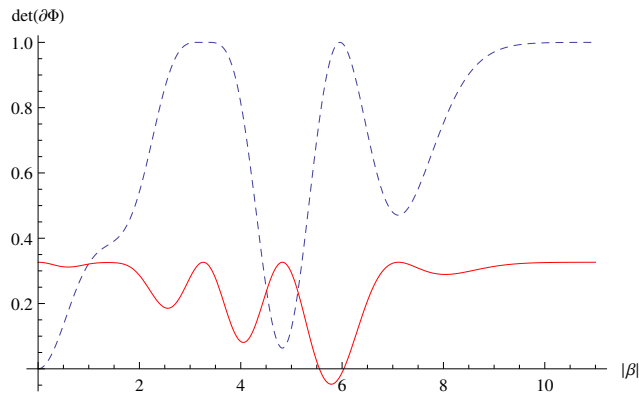

FIG. 1: Solid line shows the radial course of the GBM determinant (19). It proves nonclassicality for $|\beta| \approx 5.8$. Both, the Bochner determinant in (13) (dashed line) and the MoM determinant in (14) (the value of solid line at $|\beta| = 0$), are non-negative for this state.

## III. SAMPLING FORMULAS

We will rewrite the CF as an average of the quadrature distribution for any phase. This renders it possible to formulate a sampling approach for the GBM.

Under the assumption that all moments exist, we can expand the CF as

$$\Phi(\beta) = \langle : e^{\beta \hat{a}^\dagger - \beta^* \hat{a}} : \rangle = \sum_{k,l=0}^{\infty} \frac{\beta^k (-\beta^*)^l}{k!l!} \langle \hat{a}^{\dagger k} \hat{a}^l \rangle. \tag{20}$$

The normally ordered moments $\langle \hat{a}^{\dagger k} \hat{a}^l \rangle$ can be reconstructed from the quadrature distribution $p(x,\varphi)$ [5]:

$$\langle \hat{a}^{\dagger m} \hat{a}^n \rangle = \int_0^\pi d\varphi \int_{-\infty}^{+\infty} dx\, p(x,\varphi) \frac{e^{i(m-n)\varphi} m! n!}{\pi \sqrt{2^{m+n}} (m+n)!} H_{m+n}\left(\frac{x}{\sqrt{2}}\right), \tag{21}$$

where $H_k$ is the $k$th Hermite polynomial. Inserting Eq. (21) into Eq. (20) yields:

$$\Phi(\beta) = \int_0^\pi d\varphi \int_{-\infty}^{+\infty} dx\, \frac{p(x,\varphi)}{\pi} \sum_{k,l=0}^{\infty} \frac{\beta^k (-\beta^*)^l e^{i(k-l)\varphi}}{\sqrt{2^{k+l}} (k+l)!} H_{k+l}\left(\frac{x}{\sqrt{2}}\right). \tag{22}$$

Now, the latter sum may be rewritten ($\gamma = \beta e^{i\varphi}$):

$$\sum_{k,l=0}^{\infty} \frac{\beta^k (-\beta^*)^l e^{i(k-l)\varphi}}{\sqrt{2^{k+l}} (k+l)!} H_{k+l}\left(\frac{x}{\sqrt{2}}\right) = \sum_{p=0}^{\infty} \frac{(-\gamma)^p}{\sqrt{2^p} p!} H_p\left(\frac{x}{\sqrt{2}}\right) \sum_{q=0}^{p} \left[-\frac{\gamma}{\gamma^*}\right]^q = \sum_{p=0}^{\infty} \frac{(-\gamma)^p}{\sqrt{2^p} p!} H_p\left(\frac{x}{\sqrt{2}}\right) \frac{1-(-\gamma/\gamma^*)^{p+1}}{1-(-\gamma/\gamma^*)}. \tag{23}$$

The remaining series can be directly calculated with the generating function of Hermite polynomials [6]:

$$\sum_{n=0}^{\infty} \frac{t^n}{n!} H_n(x) = e^{2xt - t^2}. \tag{24}$$

Inserting this relation into the CF, it can be expressed with quadrature distribution $p(x,\varphi)$ and $\gamma = \beta e^{i\varphi}$ as

$$\Phi(\beta) = \int_0^\pi d\varphi \int_{-\infty}^{+\infty} dx\, \frac{p(x,\varphi)}{\pi} \frac{\gamma e^{x\gamma - \gamma^2/2} + \gamma^* e^{-x\gamma^* - \gamma^{*2}/2}}{\gamma + \gamma^*}. \tag{25}$$

For the proof of the integrand to be analytical, we consider the expansion ($\gamma = \gamma_R + i\gamma_I$):

$$f(\gamma) = \frac{\gamma e^{x\gamma - \gamma^2/2} + \gamma^* e^{-x\gamma^* - \gamma^{*2}/2}}{\gamma + \gamma^*} = \partial_x \frac{e^{x\gamma - \gamma^2/2} - e^{-x\gamma^* - \gamma^{*2}/2}}{\gamma + \gamma^*} \tag{26}$$

$$= \partial_x e^{ix\gamma_I - \gamma_R^2/2 + \gamma_I^2/2} \frac{\sinh[\gamma_R(x - i\gamma_I)]}{\gamma_R} = \partial_x e^{ix\gamma_I - \gamma_R^2/2 + \gamma_I^2/2} (x - i\gamma_I) \sum_{k=0}^{\infty} \frac{[\gamma_R(x - i\gamma_I)]^{2k}}{(2k+1)!}. \tag{27}$$

Note that the remaining series converges absolutely:

$$\sum_{k=0}^{\infty} \frac{|\gamma_R(x - i\gamma_I)|^{2k}}{(2k+1)!} \leq \sum_{k=0}^{\infty} \frac{|\gamma_R(x - i\gamma_I)|^{2k}}{(2k)!} = \cosh|\gamma_R(x - i\gamma_I)|. \tag{28}$$

The power series of the integrand converges for all $\gamma$, so the integrand is an analytical function.

Let us now derive such a representation for derivatives of the CF. Using Eq. (25) with $\gamma = \beta e^{i\varphi}$ we need to replace:

$$\partial_\beta^m \partial_{\beta^*}^n = e^{i(m-n)\varphi} \partial_\gamma^m \partial_{\gamma^*}^n. \tag{29}$$

Now, arbitrary derivatives can be computed:

$$\partial_\beta^m \partial_{\beta^*}^n \Phi(\beta) = \int_0^\pi d\varphi \int_{-\infty}^{+\infty} dx\, \frac{p(x,\varphi)}{\pi} e^{i(m-n)\varphi} \partial_\gamma^m \partial_{\gamma^*}^n \left[\frac{\gamma}{\gamma + \gamma^*} e^{x\gamma - \gamma^2/2} + \frac{\gamma^*}{\gamma + \gamma^*} e^{-x\gamma^* - \gamma^{*2}/2}\right]$$

$$= \int_0^\pi d\varphi \int_{-\infty}^{+\infty} dx\, \frac{p(x,\varphi)}{\pi} e^{i(m-n)\varphi} \left[\partial_\gamma^m \frac{(-1)^n n!}{(\gamma + \gamma^*)^{n+1}} \gamma e^{x\gamma - \gamma^2/2} + \partial_{\gamma^*}^n \frac{(-1)^m m!}{(\gamma + \gamma^*)^{m+1}} \gamma^* e^{-x\gamma^* - \gamma^{*2}/2}\right]$$

$$= \int_0^\pi d\varphi \int_{-\infty}^{+\infty} dx\, \frac{p(x,\varphi)}{\pi} e^{i(m-n)\varphi} \left[m \partial_\gamma^{m-1} \frac{(-1)^n n!}{(\gamma + \gamma^*)^{n+1}} e^{x\gamma - \gamma^2/2} + \gamma \partial_\gamma^m \frac{(-1)^n n!}{(\gamma + \gamma^*)^{n+1}} e^{x\gamma - \gamma^2/2}\right.$$

$$\left. + n \partial_{\gamma^*}^{n-1} \frac{(-1)^m m!}{(\gamma + \gamma^*)^{m+1}} e^{-x\gamma^* - \gamma^{*2}/2} + \gamma^* \partial_{\gamma^*}^n \frac{(-1)^m m!}{(\gamma + \gamma^*)^{m+1}} e^{-x\gamma^* - \gamma^{*2}/2}\right]. \tag{30}$$

Using the definition of Hermite polynomials, $\mathrm{H}_n(z/\sqrt{2}) = (-\sqrt{2})^n e^{z^2/2} \frac{d^n}{dz^n} e^{-z^2/2}$, we define the pattern functions:

$$D_q^r(x,\gamma) = \sum_{k_1+k_2+k_3=r} \frac{r!}{k_1!k_2!k_3!} \frac{(-1)^{q+k_1+k_3}(q+k_1)! 2^{-k_3/2}}{(\gamma+\gamma^*)^{q+k_1+1}} x^{k_2} e^{x\gamma-\gamma^2/2} \mathrm{H}_{k_3}\left(\frac{\gamma}{\sqrt{2}}\right), \qquad (31)$$

which has to be set to zero $D_q^r \equiv 0$ for $q<0$ or $r<0$. After evaluating all terms of derivatives the derivatives of the CF reads as

$$\partial_\beta^m \partial_{\beta^*}^n \Phi(\beta) = \int_0^\pi d\varphi \int_{-\infty}^{+\infty} dx \, \frac{p(x,\varphi)}{\pi} e^{i(m-n)\varphi} \left[ m D_n^{m-1}(x, \beta e^{i\varphi}) + \beta e^{i\varphi} D_n^m(x, \beta e^{i\varphi}) \right.$$
$$\left. + n D_m^{n-1}(-x, \beta^* e^{-i\varphi}) + \beta^* e^{-i\varphi} D_m^n(-x, \beta^* e^{-i\varphi}) \right]. \qquad (32)$$

Using the standard sampling approach for quadrature-phase data points $(x_j, \varphi_j)_{j=1}^M$, we get the estimate

$$\partial_\beta^m \partial_{\beta^*}^n \Phi(\beta) \approx \frac{1}{M} \sum_{j=1}^M e^{i(m-n)\varphi_j} \left[ m D_n^{m-1}(x_j, \beta e^{i\varphi_j}) + \beta e^{i\varphi_j} D_n^m(x_j, \beta e^{i\varphi_j}) \right.$$
$$\left. + n D_m^{n-1}(-x_j, \beta^* e^{-i\varphi_j}) + \beta^* e^{-i\varphi_j} D_m^n(-x_j, \beta^* e^{-i\varphi_j}) \right]. \qquad (33)$$

## IV. EXPERIMENTAL RESULTS

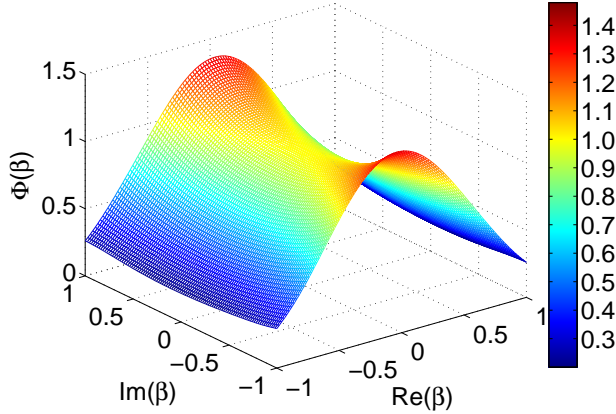

FIG. 2: Using Eq. (33) for $m=n=0$, the sampled CF, $\mathrm{Re}[\Phi(\beta)]$, is shown. The imaginary part is zero within the sampling error, $\mathrm{Im}[\Phi(\beta)] \approx 0$. It is important to stress that the behavior along the imaginary axis ($\mathrm{Re}(\beta) \approx 0$) has an inverse Gaussian slope.

The sampled CF of the measured state is shown in Fig. 2. This squeezed vacuum state is theoretically described by

$$\Phi_{\mathrm{sv}}(\beta) = e^{-\mathrm{Re}(\beta e^{i\varphi})^2 V_a/2 + \mathrm{Im}(\beta e^{i\varphi})^2 V_s/2 + |\beta|^2/2} \in \mathbb{R}, \qquad (34)$$

where $V_a$ and $V_s$ are squeezed and anti-squeezed variances, respectively, and $\varphi$ describes a rotation of the principal-axes (squeezed and anti-squeezed quadratures) in phase-space. The experimentally obtained variances are $V_a = 4.086$ (or $6.11\,\mathrm{dB}$ anti-squeezing) and $V_s = 0.386$ (or $-4.13\,\mathrm{dB}$ squeezing). Note that for vacuum holds $V_a = V_s = 1$. For $N=3$ and the parameters $(m_1, m_2, m_3) = (0,1,0)$, $(n_1, n_2, n_3) = (0,0,1)$, and $(\beta_1, \beta_2, \beta_3) = (\beta, 0, \beta)$, the GBM in (4) reads as

$$\partial\Phi = \begin{pmatrix} \Phi(0) & \partial_\beta \Phi(\beta) & \partial_{\beta^*} \Phi(0) \\ -\partial_{\beta^*} \Phi(-\beta) & -\partial_\beta \partial_{\beta^*} \Phi(0) & -\partial_{\beta^*}^2 \Phi(-\beta) \\ -\partial_\beta \Phi(0) & -\partial_\beta^2 \Phi(\beta) & -\partial_\beta \partial_{\beta^*} \Phi(0) \end{pmatrix}. \qquad (35)$$

Using (10) the determinant for $\beta = 0$ can be rewritten as:

$$\det(\partial\Phi)\Big|_{\beta=0} = \det\begin{pmatrix} 1 & \langle\hat{a}^\dagger\rangle & \langle\hat{a}\rangle \\ \langle\hat{a}\rangle & \langle\hat{a}^\dagger\hat{a}\rangle & \langle\hat{a}^2\rangle \\ \langle\hat{a}^\dagger\rangle & \langle\hat{a}^{\dagger 2}\rangle & \langle\hat{a}^\dagger\hat{a}\rangle \end{pmatrix} = \frac{1}{4}\langle:[\Delta\hat{x}(\varphi_{\min})]^2:\rangle\langle:[\Delta\hat{x}(\varphi_{\max})]^2:\rangle = \frac{1}{4}(V_a - 1)(V_s - 1), \qquad (36)$$

being known from quadrature squeezing condition [1]. The last equality holds only for Gaussian states, which are fully characterized by two variances $V_a$ and $V_s$. The sampled minor at the origin is $\det(\partial\Phi)|_{\beta=0} = -0.469 \pm 0.007$.

---



\end{document}